\documentstyle[preprint,aps]{revtex}
\renewcommand{\baselinestretch}{1.5}

\begin{document}
\title{ Raman continuum in high-$T_c$ superconductors}
\author{Haranath Ghosh$^1$ and Manas Sardar$^2$ \\
$^1$Institute of Physics, Sachivalaya Marg,
Bhubaneswar 751 005, India.  \\ 
$^2$Institute of theoretical Physics,
University of Regensburg, D-93040 Regensburg, Germany.}

\abstract{ 
 Experimentally, Raman continuum
for $A_g$ and $B_g$ geometry exhibits peaks far apart from each
other (about 80 to 150 $\rm cm^{-1}$) in frequency. The former is
insensitive to doping over a small range where $T_c$ does not
vary much whereas the latter shifts towards higher frequencies.
We calculate the electronic Raman scattering intensities using the
`modified spin bag model'. We show that the calculated results
have natural explanation to the observed anamolous peak
separation and peculiar doping dependence.}

\maketitle

E-mail : hng@iopb.ernet.in \\

Fax : 0674 - 481142

\thispagestyle{empty}

\twocolumn

 Raman scattering is a powerful technique to probe directly
the nature of low energy quasiparticle excitaions in superconductors.
Raman scattering experiments have been carried out\cite{1,2,3,4}
to investigate the low energy excitations ($\omega < 1000 {\rm cm}^{-1}$)
of the electronic continuum in the cuprate superconductors.
 The main puzzling features are,

(1) The Raman continuum
which is almost flat for $T >T_c $, becomes depleted
below $T_c$ at low frequencies ( $\omega < 200 {~\rm cm}^{-1}~<2\Delta_{sc}$)
and a broad peak develops in the range $250 - 600 {~\rm cm }^{-1}$.
The central frequency of this peak depends on the material studied
and the scattering geometry (that is the part of the Fermi
surface (FS) being explored), but it is developed much above the
superconducting gap threshold.
The dependence of the 
peak position on the scattering geometry shows that the superconducting  
(SC) gap is highly anisotropic.

(2) There is some residual intensity below $\omega < 2\Delta_{sc}$,
showing that scattering with quasiparticles still exists even below the 
 $T_c$. It is generally believed to be due to possible existence
of nodes on the gap function, and hence the availability of low
energy quasiparticles even below the $T_c$.

For superconductors in which the penetration depth of the incident
light is much greater than the BCS coherence length, the 
electronic Raman scattering intensity follows from the $q\approx 0$
limit only, $q$ being the momentum transfer to the quasiparicles.
The Raman scattered intensity due to the scattering by the superconducting
quasiparticles is given by,
\begin{equation}
I(0,\omega )~~=~~{1\over \pi^{2} \omega } \int_{\omega >2\vert \bigtriangleup_k
\vert} {\vert \gamma_k\vert^2 \bigtriangleup_k^2 {\rm d}^2 {\rm k}\over
(\omega^2 - 4 \bigtriangleup_k ^2)^{1/2}}\tanh ({\omega\over 4T})
\end{equation}
The integral is over the Fermi surface.

 It is observed that, for the $A_{1g}$ mode, the Raman
intensity decreases linearly with $\omega$ at lower frequencies
and extrapolates to zero at $\omega =0$. This seems to support 
a d-wave kind of SC-gap which has nodes on the Fermi surface,
and the $A_{1g}$ continuum is due to excitations of superconducting
pairs accross an anisotropic with d-wave type nodes, with
$2\bigtriangleup_{{\rm max}} =310~ {\rm cm }^{-1}$.
The surprising thing about the $A_{1g}$ continuum is that the 
peak position is independent of small doping variation and there 
is very little variation in intensity also.  
 
     But for the $B_{1g}$ mode, the peak frequency differs from
the $A_{1g}$ continuum peak by about $200~ {\rm cm}^{-1}$. Also
this peak position shifts towards higher frequencies for compounds
with lesser concentration of oxygen or rather for 
lower hole doping.

There are arguments\cite{dev}, saying that one can obtain different 
peak positions, once one takes the screening effects into account.
On the other hand, it has been argued \cite{dev,screen} that,
screening is effective more in the s-wave scattering channel, and
hence a look at the  
different structure factors for the two phonons tells us that
screening should substantially reduce
the $A_{1g}$ phonon scattered intensity and the intensity of the
$B_{1g}$ mode will not be affected much. In other words the $B_{1g}$
continuum will be stronger than the $A_{1g}$ continuum. This is
exactly opposite to what is observed experimentally \cite{8}.
Another important point worth emphasizing is that, since the $A_{1g}$
and $B_{1g}$ modes corresponds to the breathing oxygen atom
vibrations and antisymmetric out of plane vibrations of the oxygen
cage around the Cu atom respectively, one expects the $A_{1g}$ 
continuum to be sensitive to the carrier concentrations  or the oxygen
content, while the $B_{1g}$ continuum should be relatively unaffected.
This again is exactly opposite to what is experimentally observed\cite{8}.
Screening doesn't seem to be important at all.
The alternative hypothesis put forward, is that the $B_{1g}$ peak
is not associated with the superconducting gap at all, since the peaking 
in intensity at higher frequency for this mode is observed even 
slightly above $T_c$. Scattering with spin fluctuations is put forward
as a possible explanation for this, since for lower doped materials
a spin gap like feature is observed above $T_c$ and the magnitude of 
which reduces with doping.
It is argued that, if the peaking in the $B_{1g}$ mode is due to the 
superconducting gap then how could the peak position vary by 
about 20 percent, within a doping variation 
that changes $T_c$ very little? Here we explore these features
theoretically within our model.

Our model is based on the idea of Schrieffer's spin bag model \cite{9}
with coupling between the planes explicitly introduced to analyze two
layer systems. At half filling due to the square planar nature
of the Fermi surface ( nesting ) and intermediate inplane
correlation between the Cu spins leads to SDW insulating phase
with a fully gapped FS. With doping, nesting is lost near the
$M$ points and hence the SDW gap vanishes in these directions while 
surviving in other directions. This effect is enhanced with the
introduction of single particle tunneling between the planes.
The pairing interaction between the low energy SDW
quasiparticles in the gapless regions is mediated by the 
quanta of fluctuations of the amplitude and phase modes of the
SDW gap in the surviving regions \cite{10}. 
Notable difference with earlier treatments is that the
conduction and 
valence bands touch each other in the regions 
 where the nesting is lost and thereby the SDW gap
$G$ is assumed to be zero in these regions while solving the
self-consistent SDW gap equation. 
The coupling between the SDW quasiparticles and the fluctuations
of the amplitude and phase (i.e, collective) modes of the SDW
state, will give rise to new kind of electron-amplitudon
(phason) interaction. Such interaction in a second
order perturbation theory gives rise to an effective pairing
interaction is the essence of the modified spin bag mechanism \cite{10}.
The superconducting gap equation from our model is given below,
\begin{eqnarray}
\Delta_{sc}(k) && = 
\sum_{k^{\prime}}[\lambda_{1}+\lambda_2{(\epsilon_k-\mu)
(\epsilon_{k^{\prime}}-\mu)
\over E_{k}E_{k^{\prime}}} \nonumber \\
& &
-{G(k)G(k^\prime)\over E_{k}E_{k^\prime}}
]
(\Delta_{sc}(k^{\prime})/e_{k^{\prime}}) \nonumber \\  
& &  
\times \tanh(\beta
e_{k^{\prime}}/2) 
\end{eqnarray}
where, $\lambda_{1(2)}=\Omega_{AM}U^2/[(E_k\mp
E_{k^{\prime}})^2-\Omega_{AM}^2]$, and $\Omega_{AM}$ 
being the maximum frequency of the SDW gap fluctuation (
amplitudon) given by, $\Omega_{AM}=2G_{max}$. Where
$e_k=\sqrt{E_{k}^2 + \Delta_{sc}^2}$ 
and $E_k=[(\epsilon_k -\mu)^2$ $+G^2]^{1/2}$ are respectively
the SC and SDW quasi particle energies. It is clear that
the SC-gap will peak up to its maximum for the ($k_x, k_y$)
points where G(k) vanishes and hence the two gaps (SDW and SC)
will have complementary nature.

We solved both the SDW and superconducting gap equations
numerically for different filling factors. Parameters chosen
are, $t=0.3$ eV, $U=2.0 $ eV, pairing interaction cutoff for the
SDW gap equation to be 75 meV. With these values and for chemical
potential $\mu=-250$ meV we get an SDW transition temperature
$T_{sdw}$ of $100^o$K, $t_{\perp}$ is taken to be 0.05 eV. 
For the superconducting gap equation the pairing cutoff is $-G$
to $G$ (the maximum SDW gap). We choose, $\lambda_1=50$ 
and $\lambda_2=40$ meV to get a $T_c$ of $85^o$K for $\mu=-250 $ meV.
$T_c$ decreases by 7 degrees when the chemical potential is
varied upto -200 meV (that is for a doping concentration of 0.14
to 0.12, measured from half filling). 
The self consistent gap equatoion is solved numerically for 3
different doping concentrations. 
The main feature of the gap is that, it is larger near the $M$
points than near the $X$ points by 3-4 meV. Also the gap near 
the $M$ points falls slower with temperature than the gaps near
the $X$ points. 
Of course we do not have
any gap nodes and the gap values near the $X$ points is still
substantial ( 10-12 meV) for such small doping concentrations.
So we do not get any Raman intensity in the low frequency
region. We discuss the results of our numerical calculations
below.

For the $B_{1g}$ mode, the structure factor $\gamma_k$
also is maximum wherever the gap is maximum, and most of the contribution
to the scattered intensity in this channel comes from quasiparticles
in these regions.
On the other hand, for the $A_{1g}$, the structure factor is large
and more or less flat in almost all regions in the Brillouin zone
and falls to low values at the places where the gap value 
is maximum. In other words most of the scattered Raman intensity in this
channel are from the region where the gap value is small.

\setlength{\unitlength}{0.240900pt}
\ifx\plotpoint\undefined\newsavebox{\plotpoint}\fi
\sbox{\plotpoint}{\rule[-0.200pt]{0.400pt}{0.400pt}}%
\begin{picture}(825,584)(0,0)
\font\gnuplot=cmr10 at 12pt
\gnuplot
\sbox{\plotpoint}{\rule[-0.200pt]{0.400pt}{0.400pt}}%
\put(120.0,31.0){\rule[-0.200pt]{159.235pt}{0.400pt}}
\put(120.0,31.0){\rule[-0.200pt]{4.818pt}{0.400pt}}
\put(108,31){\makebox(0,0)[r]{0}}
\put(761.0,31.0){\rule[-0.200pt]{4.818pt}{0.400pt}}
\put(120.0,161.0){\rule[-0.200pt]{4.818pt}{0.400pt}}
\put(108,161){\makebox(0,0)[r]{0.5}}
\put(761.0,161.0){\rule[-0.200pt]{4.818pt}{0.400pt}}
\put(120.0,291.0){\rule[-0.200pt]{4.818pt}{0.400pt}}
\put(108,291){\makebox(0,0)[r]{1}}
\put(761.0,291.0){\rule[-0.200pt]{4.818pt}{0.400pt}}
\put(120.0,421.0){\rule[-0.200pt]{4.818pt}{0.400pt}}
\put(108,421){\makebox(0,0)[r]{1.5}}
\put(761.0,421.0){\rule[-0.200pt]{4.818pt}{0.400pt}}
\put(120.0,551.0){\rule[-0.200pt]{4.818pt}{0.400pt}}
\put(108,551){\makebox(0,0)[r]{2}}
\put(761.0,551.0){\rule[-0.200pt]{4.818pt}{0.400pt}}
\put(206.0,31.0){\rule[-0.200pt]{0.400pt}{4.818pt}}
\put(206,19){\makebox(0,0){40}}
\put(206.0,557.0){\rule[-0.200pt]{0.400pt}{4.818pt}}
\put(350.0,31.0){\rule[-0.200pt]{0.400pt}{4.818pt}}
\put(350,19){\makebox(0,0){45}}
\put(350.0,557.0){\rule[-0.200pt]{0.400pt}{4.818pt}}
\put(494.0,31.0){\rule[-0.200pt]{0.400pt}{4.818pt}}
\put(494,19){\makebox(0,0){50}}
\put(494.0,557.0){\rule[-0.200pt]{0.400pt}{4.818pt}}
\put(637.0,31.0){\rule[-0.200pt]{0.400pt}{4.818pt}}
\put(637,19){\makebox(0,0){55}}
\put(637.0,557.0){\rule[-0.200pt]{0.400pt}{4.818pt}}
\put(781.0,31.0){\rule[-0.200pt]{0.400pt}{4.818pt}}
\put(781,19){\makebox(0,0){60}}
\put(781.0,557.0){\rule[-0.200pt]{0.400pt}{4.818pt}}
\put(120.0,31.0){\rule[-0.200pt]{159.235pt}{0.400pt}}
\put(781.0,31.0){\rule[-0.200pt]{0.400pt}{131.531pt}}
\put(120.0,577.0){\rule[-0.200pt]{159.235pt}{0.400pt}}
\put(12,304){\makebox(0,0){$I (0, \omega)$}}
\put(450,-37){\makebox(0,0){Frequency ($ \omega$)}}
\put(451,541){\makebox(0,0)[l]{$\mu =-250$ {\small ---} }}
\put(514,421){\makebox(0,0)[l]{$ =-225$  {\small ...}}}
\put(514,291){\makebox(0,0)[l]{$ =-200$ {\bf \small ---}}}
\put(120.0,31.0){\rule[-0.200pt]{0.400pt}{131.531pt}}
\put(140,30.67){\rule{3.373pt}{0.400pt}}
\multiput(140.00,30.17)(7.000,1.000){2}{\rule{1.686pt}{0.400pt}}
\put(154,32.17){\rule{3.100pt}{0.400pt}}
\multiput(154.00,31.17)(8.566,2.000){2}{\rule{1.550pt}{0.400pt}}
\put(169,32.17){\rule{2.900pt}{0.400pt}}
\multiput(169.00,33.17)(7.981,-2.000){2}{\rule{1.450pt}{0.400pt}}
\put(183,31.67){\rule{3.614pt}{0.400pt}}
\multiput(183.00,31.17)(7.500,1.000){2}{\rule{1.807pt}{0.400pt}}
\put(198,32.67){\rule{3.373pt}{0.400pt}}
\multiput(198.00,32.17)(7.000,1.000){2}{\rule{1.686pt}{0.400pt}}
\put(212,34.17){\rule{2.900pt}{0.400pt}}
\multiput(212.00,33.17)(7.981,2.000){2}{\rule{1.450pt}{0.400pt}}
\multiput(226.00,36.61)(3.141,0.447){3}{\rule{2.100pt}{0.108pt}}
\multiput(226.00,35.17)(10.641,3.000){2}{\rule{1.050pt}{0.400pt}}
\multiput(241.00,39.58)(0.536,0.493){23}{\rule{0.531pt}{0.119pt}}
\multiput(241.00,38.17)(12.898,13.000){2}{\rule{0.265pt}{0.400pt}}
\multiput(255.00,52.59)(0.786,0.489){15}{\rule{0.722pt}{0.118pt}}
\multiput(255.00,51.17)(12.501,9.000){2}{\rule{0.361pt}{0.400pt}}
\multiput(269.58,61.00)(0.494,0.942){27}{\rule{0.119pt}{0.847pt}}
\multiput(268.17,61.00)(15.000,26.243){2}{\rule{0.400pt}{0.423pt}}
\multiput(284.58,89.00)(0.494,7.475){25}{\rule{0.119pt}{5.929pt}}
\multiput(283.17,89.00)(14.000,191.695){2}{\rule{0.400pt}{2.964pt}}
\multiput(298.58,293.00)(0.494,6.551){27}{\rule{0.119pt}{5.220pt}}
\multiput(297.17,293.00)(15.000,181.166){2}{\rule{0.400pt}{2.610pt}}
\multiput(313.58,485.00)(0.494,2.590){25}{\rule{0.119pt}{2.129pt}}
\multiput(312.17,485.00)(14.000,66.582){2}{\rule{0.400pt}{1.064pt}}
\multiput(327.58,553.57)(0.494,-0.607){25}{\rule{0.119pt}{0.586pt}}
\multiput(326.17,554.78)(14.000,-15.784){2}{\rule{0.400pt}{0.293pt}}
\multiput(341.00,537.92)(0.534,-0.494){25}{\rule{0.529pt}{0.119pt}}
\multiput(341.00,538.17)(13.903,-14.000){2}{\rule{0.264pt}{0.400pt}}
\multiput(356.58,513.44)(0.494,-3.435){25}{\rule{0.119pt}{2.786pt}}
\multiput(355.17,519.22)(14.000,-88.218){2}{\rule{0.400pt}{1.393pt}}
\multiput(370.58,395.12)(0.494,-10.963){25}{\rule{0.119pt}{8.643pt}}
\multiput(369.17,413.06)(14.000,-281.061){2}{\rule{0.400pt}{4.321pt}}
\multiput(384.58,125.61)(0.494,-1.831){27}{\rule{0.119pt}{1.540pt}}
\multiput(383.17,128.80)(15.000,-50.804){2}{\rule{0.400pt}{0.770pt}}
\multiput(399.58,73.67)(0.494,-1.195){25}{\rule{0.119pt}{1.043pt}}
\multiput(398.17,75.84)(14.000,-30.835){2}{\rule{0.400pt}{0.521pt}}
\multiput(413.58,45.00)(0.494,0.737){27}{\rule{0.119pt}{0.687pt}}
\multiput(412.17,45.00)(15.000,20.575){2}{\rule{0.400pt}{0.343pt}}
\multiput(428.00,65.94)(1.943,-0.468){5}{\rule{1.500pt}{0.113pt}}
\multiput(428.00,66.17)(10.887,-4.000){2}{\rule{0.750pt}{0.400pt}}
\multiput(442.00,63.59)(0.786,0.489){15}{\rule{0.722pt}{0.118pt}}
\multiput(442.00,62.17)(12.501,9.000){2}{\rule{0.361pt}{0.400pt}}
\multiput(456.58,68.71)(0.494,-0.873){27}{\rule{0.119pt}{0.793pt}}
\multiput(455.17,70.35)(15.000,-24.353){2}{\rule{0.400pt}{0.397pt}}
\multiput(471.00,44.93)(1.214,-0.482){9}{\rule{1.033pt}{0.116pt}}
\multiput(471.00,45.17)(11.855,-6.000){2}{\rule{0.517pt}{0.400pt}}
\multiput(485.58,40.00)(0.494,1.709){25}{\rule{0.119pt}{1.443pt}}
\multiput(484.17,40.00)(14.000,44.005){2}{\rule{0.400pt}{0.721pt}}
\multiput(499.00,87.59)(1.103,0.485){11}{\rule{0.957pt}{0.117pt}}
\multiput(499.00,86.17)(13.013,7.000){2}{\rule{0.479pt}{0.400pt}}
\multiput(514.58,87.42)(0.494,-1.892){25}{\rule{0.119pt}{1.586pt}}
\multiput(513.17,90.71)(14.000,-48.709){2}{\rule{0.400pt}{0.793pt}}
\put(528,41.67){\rule{3.373pt}{0.400pt}}
\multiput(528.00,41.17)(7.000,1.000){2}{\rule{1.686pt}{0.400pt}}
\multiput(542.00,43.59)(1.304,0.482){9}{\rule{1.100pt}{0.116pt}}
\multiput(542.00,42.17)(12.717,6.000){2}{\rule{0.550pt}{0.400pt}}
\multiput(557.00,49.58)(0.582,0.492){21}{\rule{0.567pt}{0.119pt}}
\multiput(557.00,48.17)(12.824,12.000){2}{\rule{0.283pt}{0.400pt}}
\multiput(571.58,57.82)(0.494,-0.839){27}{\rule{0.119pt}{0.767pt}}
\multiput(570.17,59.41)(15.000,-23.409){2}{\rule{0.400pt}{0.383pt}}
\multiput(586.58,36.00)(0.494,0.717){25}{\rule{0.119pt}{0.671pt}}
\multiput(585.17,36.00)(14.000,18.606){2}{\rule{0.400pt}{0.336pt}}
\multiput(600.00,56.58)(0.637,0.492){19}{\rule{0.609pt}{0.118pt}}
\multiput(600.00,55.17)(12.736,11.000){2}{\rule{0.305pt}{0.400pt}}
\multiput(614.58,63.04)(0.494,-1.079){27}{\rule{0.119pt}{0.953pt}}
\multiput(613.17,65.02)(15.000,-30.021){2}{\rule{0.400pt}{0.477pt}}
\put(120.0,31.0){\rule[-0.200pt]{4.818pt}{0.400pt}}
\multiput(643.58,35.00)(0.494,0.607){25}{\rule{0.119pt}{0.586pt}}
\multiput(642.17,35.00)(14.000,15.784){2}{\rule{0.400pt}{0.293pt}}
\multiput(657.58,52.00)(0.494,0.668){27}{\rule{0.119pt}{0.633pt}}
\multiput(656.17,52.00)(15.000,18.685){2}{\rule{0.400pt}{0.317pt}}
\multiput(672.58,67.32)(0.494,-1.305){25}{\rule{0.119pt}{1.129pt}}
\multiput(671.17,69.66)(14.000,-33.658){2}{\rule{0.400pt}{0.564pt}}
\multiput(686.00,36.58)(0.625,0.492){21}{\rule{0.600pt}{0.119pt}}
\multiput(686.00,35.17)(13.755,12.000){2}{\rule{0.300pt}{0.400pt}}
\multiput(701.00,46.93)(1.026,-0.485){11}{\rule{0.900pt}{0.117pt}}
\multiput(701.00,47.17)(12.132,-7.000){2}{\rule{0.450pt}{0.400pt}}
\put(629.0,35.0){\rule[-0.200pt]{3.373pt}{0.400pt}}
\multiput(729.00,41.58)(0.684,0.492){19}{\rule{0.645pt}{0.118pt}}
\multiput(729.00,40.17)(13.660,11.000){2}{\rule{0.323pt}{0.400pt}}
\multiput(744.00,50.94)(1.943,-0.468){5}{\rule{1.500pt}{0.113pt}}
\multiput(744.00,51.17)(10.887,-4.000){2}{\rule{0.750pt}{0.400pt}}
\multiput(758.58,48.00)(0.494,0.570){25}{\rule{0.119pt}{0.557pt}}
\multiput(757.17,48.00)(14.000,14.844){2}{\rule{0.400pt}{0.279pt}}
\multiput(772.59,60.45)(0.489,-0.961){15}{\rule{0.118pt}{0.856pt}}
\multiput(771.17,62.22)(9.000,-15.224){2}{\rule{0.400pt}{0.428pt}}
\put(715.0,41.0){\rule[-0.200pt]{3.373pt}{0.400pt}}
\put(120.00,31.00){\usebox{\plotpoint}}
\put(140.73,31.65){\usebox{\plotpoint}}
\multiput(146,32)(20.703,1.479){0}{\usebox{\plotpoint}}
\put(161.44,32.90){\usebox{\plotpoint}}
\put(182.16,32.00){\usebox{\plotpoint}}
\put(202.88,32.99){\usebox{\plotpoint}}
\multiput(203,33)(20.573,2.743){0}{\usebox{\plotpoint}}
\put(223.45,35.78){\usebox{\plotpoint}}
\put(241.86,44.04){\usebox{\plotpoint}}
\put(260.55,52.82){\usebox{\plotpoint}}
\put(272.24,69.87){\usebox{\plotpoint}}
\multiput(275,74)(2.052,20.654){7}{\usebox{\plotpoint}}
\multiput(290,225)(2.036,20.655){7}{\usebox{\plotpoint}}
\multiput(304,367)(5.301,20.067){3}{\usebox{\plotpoint}}
\multiput(318,420)(16.207,-12.966){0}{\usebox{\plotpoint}}
\put(333.13,407.89){\usebox{\plotpoint}}
\multiput(347,397)(4.070,-20.352){4}{\usebox{\plotpoint}}
\multiput(361,327)(1.406,-20.708){10}{\usebox{\plotpoint}}
\multiput(376,106)(12.743,-16.383){2}{\usebox{\plotpoint}}
\put(399.75,71.10){\usebox{\plotpoint}}
\put(414.65,58.55){\usebox{\plotpoint}}
\put(429.26,44.54){\usebox{\plotpoint}}
\put(447.59,37.08){\usebox{\plotpoint}}
\multiput(448,37)(19.546,6.981){0}{\usebox{\plotpoint}}
\put(465.13,46.48){\usebox{\plotpoint}}
\multiput(476,62)(12.064,-16.889){2}{\usebox{\plotpoint}}
\multiput(491,41)(19.077,8.176){0}{\usebox{\plotpoint}}
\put(506.98,45.87){\usebox{\plotpoint}}
\put(525.90,39.46){\usebox{\plotpoint}}
\put(543.93,47.81){\usebox{\plotpoint}}
\put(558.66,39.63){\usebox{\plotpoint}}
\put(570.74,47.16){\usebox{\plotpoint}}
\put(582.98,50.16){\usebox{\plotpoint}}
\put(598.57,45.04){\usebox{\plotpoint}}
\put(614.42,39.98){\usebox{\plotpoint}}
\put(629.89,41.77){\usebox{\plotpoint}}
\multiput(634,45)(20.573,2.743){0}{\usebox{\plotpoint}}
\put(649.36,46.84){\usebox{\plotpoint}}
\put(668.92,41.00){\usebox{\plotpoint}}
\put(688.14,46.79){\usebox{\plotpoint}}
\put(703.53,37.47){\usebox{\plotpoint}}
\multiput(706,35)(20.756,0.000){0}{\usebox{\plotpoint}}
\put(722.78,36.40){\usebox{\plotpoint}}
\put(739.79,43.95){\usebox{\plotpoint}}
\put(756.59,47.59){\usebox{\plotpoint}}
\put(769.89,46.58){\usebox{\plotpoint}}
\multiput(778,35)(20.756,0.000){0}{\usebox{\plotpoint}}
\put(781,35){\usebox{\plotpoint}}
\sbox{\plotpoint}{\rule[-0.400pt]{0.800pt}{0.800pt}}%
\put(123,29.84){\rule{3.373pt}{0.800pt}}
\multiput(123.00,29.34)(7.000,1.000){2}{\rule{1.686pt}{0.800pt}}
\put(137,30.84){\rule{3.614pt}{0.800pt}}
\multiput(137.00,30.34)(7.500,1.000){2}{\rule{1.807pt}{0.800pt}}
\put(152,30.84){\rule{3.373pt}{0.800pt}}
\multiput(152.00,31.34)(7.000,-1.000){2}{\rule{1.686pt}{0.800pt}}
\put(120.0,31.0){\usebox{\plotpoint}}
\put(180,30.84){\rule{3.614pt}{0.800pt}}
\multiput(180.00,30.34)(7.500,1.000){2}{\rule{1.807pt}{0.800pt}}
\put(195,31.84){\rule{3.373pt}{0.800pt}}
\multiput(195.00,31.34)(7.000,1.000){2}{\rule{1.686pt}{0.800pt}}
\put(209,33.34){\rule{3.373pt}{0.800pt}}
\multiput(209.00,32.34)(7.000,2.000){2}{\rule{1.686pt}{0.800pt}}
\multiput(223.00,37.40)(0.863,0.516){11}{\rule{1.533pt}{0.124pt}}
\multiput(223.00,34.34)(11.817,9.000){2}{\rule{0.767pt}{0.800pt}}
\multiput(238.00,46.38)(1.936,0.560){3}{\rule{2.440pt}{0.135pt}}
\multiput(238.00,43.34)(8.936,5.000){2}{\rule{1.220pt}{0.800pt}}
\multiput(253.41,50.00)(0.508,0.635){23}{\rule{0.122pt}{1.213pt}}
\multiput(250.34,50.00)(15.000,16.482){2}{\rule{0.800pt}{0.607pt}}
\multiput(268.41,69.00)(0.509,4.955){21}{\rule{0.123pt}{7.686pt}}
\multiput(265.34,69.00)(14.000,115.048){2}{\rule{0.800pt}{3.843pt}}
\multiput(282.41,200.00)(0.509,4.650){21}{\rule{0.123pt}{7.229pt}}
\multiput(279.34,200.00)(14.000,107.997){2}{\rule{0.800pt}{3.614pt}}
\multiput(296.41,323.00)(0.508,1.590){23}{\rule{0.122pt}{2.653pt}}
\multiput(293.34,323.00)(15.000,40.493){2}{\rule{0.800pt}{1.327pt}}
\multiput(310.00,367.08)(0.639,-0.512){15}{\rule{1.218pt}{0.123pt}}
\multiput(310.00,367.34)(11.472,-11.000){2}{\rule{0.609pt}{0.800pt}}
\multiput(324.00,356.08)(0.800,-0.516){11}{\rule{1.444pt}{0.124pt}}
\multiput(324.00,356.34)(11.002,-9.000){2}{\rule{0.722pt}{0.800pt}}
\multiput(339.41,334.89)(0.508,-2.086){23}{\rule{0.122pt}{3.400pt}}
\multiput(336.34,341.94)(15.000,-52.943){2}{\rule{0.800pt}{1.700pt}}
\multiput(354.41,242.39)(0.509,-7.319){21}{\rule{0.123pt}{11.229pt}}
\multiput(351.34,265.69)(14.000,-169.695){2}{\rule{0.800pt}{5.614pt}}
\multiput(368.41,87.64)(0.508,-1.166){23}{\rule{0.122pt}{2.013pt}}
\multiput(365.34,91.82)(15.000,-29.821){2}{\rule{0.800pt}{1.007pt}}
\multiput(382.00,63.40)(0.639,0.512){15}{\rule{1.218pt}{0.123pt}}
\multiput(382.00,60.34)(11.472,11.000){2}{\rule{0.609pt}{0.800pt}}
\multiput(396.00,71.08)(0.581,-0.511){17}{\rule{1.133pt}{0.123pt}}
\multiput(396.00,71.34)(11.648,-12.000){2}{\rule{0.567pt}{0.800pt}}
\multiput(410.00,59.08)(0.626,-0.511){17}{\rule{1.200pt}{0.123pt}}
\multiput(410.00,59.34)(12.509,-12.000){2}{\rule{0.600pt}{0.800pt}}
\multiput(425.00,47.08)(0.533,-0.509){19}{\rule{1.062pt}{0.123pt}}
\multiput(425.00,47.34)(11.797,-13.000){2}{\rule{0.531pt}{0.800pt}}
\multiput(439.00,37.38)(1.936,0.560){3}{\rule{2.440pt}{0.135pt}}
\multiput(439.00,34.34)(8.936,5.000){2}{\rule{1.220pt}{0.800pt}}
\multiput(454.41,41.00)(0.508,0.670){23}{\rule{0.122pt}{1.267pt}}
\multiput(451.34,41.00)(15.000,17.371){2}{\rule{0.800pt}{0.633pt}}
\multiput(469.41,54.95)(0.509,-0.798){21}{\rule{0.123pt}{1.457pt}}
\multiput(466.34,57.98)(14.000,-18.976){2}{\rule{0.800pt}{0.729pt}}
\multiput(482.00,40.39)(1.355,0.536){5}{\rule{2.067pt}{0.129pt}}
\multiput(482.00,37.34)(9.711,6.000){2}{\rule{1.033pt}{0.800pt}}
\put(496,45.34){\rule{3.200pt}{0.800pt}}
\multiput(496.00,43.34)(8.358,4.000){2}{\rule{1.600pt}{0.800pt}}
\multiput(511.00,47.08)(0.710,-0.514){13}{\rule{1.320pt}{0.124pt}}
\multiput(511.00,47.34)(11.260,-10.000){2}{\rule{0.660pt}{0.800pt}}
\put(525,39.34){\rule{3.200pt}{0.800pt}}
\multiput(525.00,37.34)(8.358,4.000){2}{\rule{1.600pt}{0.800pt}}
\multiput(540.00,41.08)(0.920,-0.520){9}{\rule{1.600pt}{0.125pt}}
\multiput(540.00,41.34)(10.679,-8.000){2}{\rule{0.800pt}{0.800pt}}
\put(554,32.84){\rule{3.373pt}{0.800pt}}
\multiput(554.00,33.34)(7.000,-1.000){2}{\rule{1.686pt}{0.800pt}}
\multiput(569.41,34.00)(0.508,1.236){23}{\rule{0.122pt}{2.120pt}}
\multiput(566.34,34.00)(15.000,31.600){2}{\rule{0.800pt}{1.060pt}}
\multiput(584.41,63.48)(0.509,-0.874){21}{\rule{0.123pt}{1.571pt}}
\multiput(581.34,66.74)(14.000,-20.738){2}{\rule{0.800pt}{0.786pt}}
\multiput(597.00,44.08)(0.581,-0.511){17}{\rule{1.133pt}{0.123pt}}
\multiput(597.00,44.34)(11.648,-12.000){2}{\rule{0.567pt}{0.800pt}}
\multiput(611.00,35.40)(0.689,0.512){15}{\rule{1.291pt}{0.123pt}}
\multiput(611.00,32.34)(12.321,11.000){2}{\rule{0.645pt}{0.800pt}}
\put(166.0,32.0){\rule[-0.400pt]{3.373pt}{0.800pt}}
\multiput(640.00,43.08)(0.626,-0.511){17}{\rule{1.200pt}{0.123pt}}
\multiput(640.00,43.34)(12.509,-12.000){2}{\rule{0.600pt}{0.800pt}}
\multiput(656.41,33.00)(0.509,0.874){21}{\rule{0.123pt}{1.571pt}}
\multiput(653.34,33.00)(14.000,20.738){2}{\rule{0.800pt}{0.786pt}}
\multiput(670.41,50.48)(0.509,-0.874){21}{\rule{0.123pt}{1.571pt}}
\multiput(667.34,53.74)(14.000,-20.738){2}{\rule{0.800pt}{0.786pt}}
\put(683,31.84){\rule{3.614pt}{0.800pt}}
\multiput(683.00,31.34)(7.500,1.000){2}{\rule{1.807pt}{0.800pt}}
\put(626.0,45.0){\rule[-0.400pt]{3.373pt}{0.800pt}}
\put(712,32.84){\rule{3.373pt}{0.800pt}}
\multiput(712.00,32.34)(7.000,1.000){2}{\rule{1.686pt}{0.800pt}}
\put(726,32.84){\rule{3.614pt}{0.800pt}}
\multiput(726.00,33.34)(7.500,-1.000){2}{\rule{1.807pt}{0.800pt}}
\multiput(742.41,34.00)(0.509,0.645){21}{\rule{0.123pt}{1.229pt}}
\multiput(739.34,34.00)(14.000,15.450){2}{\rule{0.800pt}{0.614pt}}
\multiput(756.41,47.18)(0.508,-0.599){23}{\rule{0.122pt}{1.160pt}}
\multiput(753.34,49.59)(15.000,-15.592){2}{\rule{0.800pt}{0.580pt}}
\put(770,33.84){\rule{2.650pt}{0.800pt}}
\multiput(770.00,32.34)(5.500,3.000){2}{\rule{1.325pt}{0.800pt}}
\put(698.0,34.0){\rule[-0.400pt]{3.373pt}{0.800pt}}
\end{picture}

\vspace{0.5cm}

\noindent {\small {\bf Fig.1} Electronic Raman continuum intensity versus
frequency ( in meV ) in the $A_{1g}$ geometry at T= $20^o$K,
for different dopings.} \\

\setlength{\unitlength}{0.240900pt}
\ifx\plotpoint\undefined\newsavebox{\plotpoint}\fi
\sbox{\plotpoint}{\rule[-0.200pt]{0.400pt}{0.400pt}}%
\begin{picture}(825,584)(0,0)
\font\gnuplot=cmr10 at 12pt
\gnuplot
\sbox{\plotpoint}{\rule[-0.200pt]{0.400pt}{0.400pt}}%
\put(120.0,31.0){\rule[-0.200pt]{159.235pt}{0.400pt}}
\put(120.0,31.0){\rule[-0.200pt]{4.818pt}{0.400pt}}
\put(108,31){\makebox(0,0)[r]{0}}
\put(761.0,31.0){\rule[-0.200pt]{4.818pt}{0.400pt}}
\put(120.0,161.0){\rule[-0.200pt]{4.818pt}{0.400pt}}
\put(108,161){\makebox(0,0)[r]{0.5}}
\put(761.0,161.0){\rule[-0.200pt]{4.818pt}{0.400pt}}
\put(120.0,291.0){\rule[-0.200pt]{4.818pt}{0.400pt}}
\put(108,291){\makebox(0,0)[r]{1}}
\put(761.0,291.0){\rule[-0.200pt]{4.818pt}{0.400pt}}
\put(120.0,421.0){\rule[-0.200pt]{4.818pt}{0.400pt}}
\put(108,421){\makebox(0,0)[r]{1.5}}
\put(761.0,421.0){\rule[-0.200pt]{4.818pt}{0.400pt}}
\put(120.0,551.0){\rule[-0.200pt]{4.818pt}{0.400pt}}
\put(108,551){\makebox(0,0)[r]{2}}
\put(761.0,551.0){\rule[-0.200pt]{4.818pt}{0.400pt}}
\put(120.0,31.0){\rule[-0.200pt]{0.400pt}{4.818pt}}
\put(120,19){\makebox(0,0){45}}
\put(120.0,557.0){\rule[-0.200pt]{0.400pt}{4.818pt}}
\put(285.0,31.0){\rule[-0.200pt]{0.400pt}{4.818pt}}
\put(285,19){\makebox(0,0){50}}
\put(285.0,557.0){\rule[-0.200pt]{0.400pt}{4.818pt}}
\put(451.0,31.0){\rule[-0.200pt]{0.400pt}{4.818pt}}
\put(451,19){\makebox(0,0){55}}
\put(451.0,557.0){\rule[-0.200pt]{0.400pt}{4.818pt}}
\put(616.0,31.0){\rule[-0.200pt]{0.400pt}{4.818pt}}
\put(616,19){\makebox(0,0){60}}
\put(616.0,557.0){\rule[-0.200pt]{0.400pt}{4.818pt}}
\put(781.0,31.0){\rule[-0.200pt]{0.400pt}{4.818pt}}
\put(781,19){\makebox(0,0){65}}
\put(781.0,557.0){\rule[-0.200pt]{0.400pt}{4.818pt}}
\put(120.0,31.0){\rule[-0.200pt]{159.235pt}{0.400pt}}
\put(781.0,31.0){\rule[-0.200pt]{0.400pt}{131.531pt}}
\put(120.0,577.0){\rule[-0.200pt]{159.235pt}{0.400pt}}
\put(12,304){\makebox(0,0){$I (0, \omega)$}}
\put(450,-47){\makebox(0,0){Frequency ($ \omega$)}}
\put(219,525){\makebox(0,0)[l]{$\mu = -250${\small ---}}}
\put(269,421){\makebox(0,0)[l]{$= - 225${...}}}
\put(269,291){\makebox(0,0)[l]{$= -200${\bf \small ---}}}
\put(120.0,31.0){\rule[-0.200pt]{0.400pt}{131.531pt}}
\put(120,41){\usebox{\plotpoint}}
\put(120,39.67){\rule{4.095pt}{0.400pt}}
\multiput(120.00,40.17)(8.500,-1.000){2}{\rule{2.048pt}{0.400pt}}
\put(137,38.17){\rule{3.300pt}{0.400pt}}
\multiput(137.00,39.17)(9.151,-2.000){2}{\rule{1.650pt}{0.400pt}}
\put(170,36.67){\rule{3.854pt}{0.400pt}}
\multiput(170.00,37.17)(8.000,-1.000){2}{\rule{1.927pt}{0.400pt}}
\put(186,35.67){\rule{4.095pt}{0.400pt}}
\multiput(186.00,36.17)(8.500,-1.000){2}{\rule{2.048pt}{0.400pt}}
\multiput(203.00,36.58)(0.732,0.492){19}{\rule{0.682pt}{0.118pt}}
\multiput(203.00,35.17)(14.585,11.000){2}{\rule{0.341pt}{0.400pt}}
\multiput(219.00,45.93)(0.961,-0.489){15}{\rule{0.856pt}{0.118pt}}
\multiput(219.00,46.17)(15.224,-9.000){2}{\rule{0.428pt}{0.400pt}}
\multiput(236.00,38.59)(1.712,0.477){7}{\rule{1.380pt}{0.115pt}}
\multiput(236.00,37.17)(13.136,5.000){2}{\rule{0.690pt}{0.400pt}}
\multiput(252.00,41.94)(2.382,-0.468){5}{\rule{1.800pt}{0.113pt}}
\multiput(252.00,42.17)(13.264,-4.000){2}{\rule{0.900pt}{0.400pt}}
\multiput(269.00,39.60)(2.236,0.468){5}{\rule{1.700pt}{0.113pt}}
\multiput(269.00,38.17)(12.472,4.000){2}{\rule{0.850pt}{0.400pt}}
\multiput(285.00,41.94)(2.382,-0.468){5}{\rule{1.800pt}{0.113pt}}
\multiput(285.00,42.17)(13.264,-4.000){2}{\rule{0.900pt}{0.400pt}}
\multiput(302.00,39.61)(3.365,0.447){3}{\rule{2.233pt}{0.108pt}}
\multiput(302.00,38.17)(11.365,3.000){2}{\rule{1.117pt}{0.400pt}}
\multiput(318.00,40.95)(3.588,-0.447){3}{\rule{2.367pt}{0.108pt}}
\multiput(318.00,41.17)(12.088,-3.000){2}{\rule{1.183pt}{0.400pt}}
\multiput(335.00,39.58)(0.808,0.491){17}{\rule{0.740pt}{0.118pt}}
\multiput(335.00,38.17)(14.464,10.000){2}{\rule{0.370pt}{0.400pt}}
\multiput(351.00,47.92)(0.712,-0.492){21}{\rule{0.667pt}{0.119pt}}
\multiput(351.00,48.17)(15.616,-12.000){2}{\rule{0.333pt}{0.400pt}}
\put(153.0,38.0){\rule[-0.200pt]{4.095pt}{0.400pt}}
\multiput(384.00,37.59)(1.823,0.477){7}{\rule{1.460pt}{0.115pt}}
\multiput(384.00,36.17)(13.970,5.000){2}{\rule{0.730pt}{0.400pt}}
\multiput(401.00,40.93)(1.395,-0.482){9}{\rule{1.167pt}{0.116pt}}
\multiput(401.00,41.17)(13.579,-6.000){2}{\rule{0.583pt}{0.400pt}}
\multiput(417.58,36.00)(0.495,0.768){31}{\rule{0.119pt}{0.712pt}}
\multiput(416.17,36.00)(17.000,24.523){2}{\rule{0.400pt}{0.356pt}}
\multiput(434.58,59.44)(0.495,-0.648){31}{\rule{0.119pt}{0.618pt}}
\multiput(433.17,60.72)(17.000,-20.718){2}{\rule{0.400pt}{0.309pt}}
\multiput(451.00,40.58)(0.732,0.492){19}{\rule{0.682pt}{0.118pt}}
\multiput(451.00,39.17)(14.585,11.000){2}{\rule{0.341pt}{0.400pt}}
\multiput(467.58,51.00)(0.495,0.558){31}{\rule{0.119pt}{0.547pt}}
\multiput(466.17,51.00)(17.000,17.865){2}{\rule{0.400pt}{0.274pt}}
\multiput(484.58,70.00)(0.494,3.314){29}{\rule{0.119pt}{2.700pt}}
\multiput(483.17,70.00)(16.000,98.396){2}{\rule{0.400pt}{1.350pt}}
\multiput(500.58,174.00)(0.495,6.423){31}{\rule{0.119pt}{5.135pt}}
\multiput(499.17,174.00)(17.000,203.341){2}{\rule{0.400pt}{2.568pt}}
\multiput(517.58,388.00)(0.494,0.657){29}{\rule{0.119pt}{0.625pt}}
\multiput(516.17,388.00)(16.000,19.703){2}{\rule{0.400pt}{0.313pt}}
\multiput(533.58,409.00)(0.495,1.219){31}{\rule{0.119pt}{1.065pt}}
\multiput(532.17,409.00)(17.000,38.790){2}{\rule{0.400pt}{0.532pt}}
\multiput(550.58,443.57)(0.494,-1.842){29}{\rule{0.119pt}{1.550pt}}
\multiput(549.17,446.78)(16.000,-54.783){2}{\rule{0.400pt}{0.775pt}}
\multiput(566.58,384.94)(0.495,-2.031){31}{\rule{0.119pt}{1.700pt}}
\multiput(565.17,388.47)(17.000,-64.472){2}{\rule{0.400pt}{0.850pt}}
\multiput(583.58,295.36)(0.494,-8.691){29}{\rule{0.119pt}{6.900pt}}
\multiput(582.17,309.68)(16.000,-257.679){2}{\rule{0.400pt}{3.450pt}}
\multiput(599.00,50.94)(2.382,-0.468){5}{\rule{1.800pt}{0.113pt}}
\multiput(599.00,51.17)(13.264,-4.000){2}{\rule{0.900pt}{0.400pt}}
\put(368.0,37.0){\rule[-0.200pt]{3.854pt}{0.400pt}}
\multiput(632.00,46.93)(1.823,-0.477){7}{\rule{1.460pt}{0.115pt}}
\multiput(632.00,47.17)(13.970,-5.000){2}{\rule{0.730pt}{0.400pt}}
\put(616.0,48.0){\rule[-0.200pt]{3.854pt}{0.400pt}}
\put(665,41.67){\rule{4.095pt}{0.400pt}}
\multiput(665.00,42.17)(8.500,-1.000){2}{\rule{2.048pt}{0.400pt}}
\put(649.0,43.0){\rule[-0.200pt]{3.854pt}{0.400pt}}
\put(120.00,48.00){\usebox{\plotpoint}}
\put(139.04,56.27){\usebox{\plotpoint}}
\put(155.32,47.13){\usebox{\plotpoint}}
\put(174.40,41.20){\usebox{\plotpoint}}
\multiput(176,41)(20.720,-1.219){0}{\usebox{\plotpoint}}
\put(195.10,39.74){\usebox{\plotpoint}}
\put(215.75,38.00){\usebox{\plotpoint}}
\put(236.48,37.34){\usebox{\plotpoint}}
\put(257.20,36.11){\usebox{\plotpoint}}
\put(274.62,46.74){\usebox{\plotpoint}}
\multiput(275,47)(18.343,-9.711){0}{\usebox{\plotpoint}}
\put(293.01,38.32){\usebox{\plotpoint}}
\put(312.92,41.84){\usebox{\plotpoint}}
\put(333.09,41.02){\usebox{\plotpoint}}
\put(353.27,40.11){\usebox{\plotpoint}}
\put(373.62,41.93){\usebox{\plotpoint}}
\multiput(374,42)(20.440,-3.607){0}{\usebox{\plotpoint}}
\put(393.87,40.18){\usebox{\plotpoint}}
\put(412.78,43.31){\usebox{\plotpoint}}
\put(431.88,37.00){\usebox{\plotpoint}}
\put(452.10,40.47){\usebox{\plotpoint}}
\put(471.73,36.80){\usebox{\plotpoint}}
\put(483.62,51.63){\usebox{\plotpoint}}
\put(495.25,55.21){\usebox{\plotpoint}}
\put(508.26,40.87){\usebox{\plotpoint}}
\multiput(523,51)(13.840,15.468){2}{\usebox{\plotpoint}}
\multiput(540,70)(4.022,20.362){4}{\usebox{\plotpoint}}
\multiput(556,151)(2.018,20.657){8}{\usebox{\plotpoint}}
\put(579.80,333.08){\usebox{\plotpoint}}
\put(595.31,345.48){\usebox{\plotpoint}}
\multiput(606,348)(4.625,-20.234){3}{\usebox{\plotpoint}}
\multiput(622,278)(11.671,-17.163){2}{\usebox{\plotpoint}}
\multiput(639,253)(1.623,-20.692){10}{\usebox{\plotpoint}}
\put(669.06,45.69){\usebox{\plotpoint}}
\put(672,45){\usebox{\plotpoint}}
\sbox{\plotpoint}{\rule[-0.400pt]{0.800pt}{0.800pt}}%
\put(120,37.84){\rule{0.723pt}{0.800pt}}
\multiput(120.00,38.34)(1.500,-1.000){2}{\rule{0.361pt}{0.800pt}}
\multiput(123.00,40.41)(0.717,0.511){17}{\rule{1.333pt}{0.123pt}}
\multiput(123.00,37.34)(14.233,12.000){2}{\rule{0.667pt}{0.800pt}}
\multiput(140.00,49.08)(0.927,-0.516){11}{\rule{1.622pt}{0.124pt}}
\multiput(140.00,49.34)(12.633,-9.000){2}{\rule{0.811pt}{0.800pt}}
\multiput(156.00,43.41)(0.717,0.511){17}{\rule{1.333pt}{0.123pt}}
\multiput(156.00,40.34)(14.233,12.000){2}{\rule{0.667pt}{0.800pt}}
\multiput(173.00,55.40)(1.263,0.526){7}{\rule{2.029pt}{0.127pt}}
\multiput(173.00,52.34)(11.790,7.000){2}{\rule{1.014pt}{0.800pt}}
\multiput(189.00,59.09)(0.527,-0.507){25}{\rule{1.050pt}{0.122pt}}
\multiput(189.00,59.34)(14.821,-16.000){2}{\rule{0.525pt}{0.800pt}}
\put(206,41.34){\rule{3.400pt}{0.800pt}}
\multiput(206.00,43.34)(8.943,-4.000){2}{\rule{1.700pt}{0.800pt}}
\put(222,38.84){\rule{4.095pt}{0.800pt}}
\multiput(222.00,39.34)(8.500,-1.000){2}{\rule{2.048pt}{0.800pt}}
\put(239,37.34){\rule{4.095pt}{0.800pt}}
\multiput(239.00,38.34)(8.500,-2.000){2}{\rule{2.048pt}{0.800pt}}
\put(272,35.84){\rule{4.095pt}{0.800pt}}
\multiput(272.00,36.34)(8.500,-1.000){2}{\rule{2.048pt}{0.800pt}}
\put(289,34.84){\rule{3.854pt}{0.800pt}}
\multiput(289.00,35.34)(8.000,-1.000){2}{\rule{1.927pt}{0.800pt}}
\multiput(305.00,37.40)(0.788,0.512){15}{\rule{1.436pt}{0.123pt}}
\multiput(305.00,34.34)(14.019,11.000){2}{\rule{0.718pt}{0.800pt}}
\multiput(322.00,45.08)(0.927,-0.516){11}{\rule{1.622pt}{0.124pt}}
\multiput(322.00,45.34)(12.633,-9.000){2}{\rule{0.811pt}{0.800pt}}
\multiput(338.00,39.38)(2.439,0.560){3}{\rule{2.920pt}{0.135pt}}
\multiput(338.00,36.34)(10.939,5.000){2}{\rule{1.460pt}{0.800pt}}
\put(355,39.34){\rule{3.400pt}{0.800pt}}
\multiput(355.00,41.34)(8.943,-4.000){2}{\rule{1.700pt}{0.800pt}}
\put(371,39.34){\rule{3.600pt}{0.800pt}}
\multiput(371.00,37.34)(9.528,4.000){2}{\rule{1.800pt}{0.800pt}}
\put(388,39.34){\rule{3.400pt}{0.800pt}}
\multiput(388.00,41.34)(8.943,-4.000){2}{\rule{1.700pt}{0.800pt}}
\put(404,38.84){\rule{4.095pt}{0.800pt}}
\multiput(404.00,37.34)(8.500,3.000){2}{\rule{2.048pt}{0.800pt}}
\put(421,38.84){\rule{3.854pt}{0.800pt}}
\multiput(421.00,40.34)(8.000,-3.000){2}{\rule{1.927pt}{0.800pt}}
\multiput(437.00,40.40)(0.877,0.514){13}{\rule{1.560pt}{0.124pt}}
\multiput(437.00,37.34)(13.762,10.000){2}{\rule{0.780pt}{0.800pt}}
\multiput(454.00,47.08)(0.671,-0.511){17}{\rule{1.267pt}{0.123pt}}
\multiput(454.00,47.34)(13.371,-12.000){2}{\rule{0.633pt}{0.800pt}}
\put(256.0,38.0){\rule[-0.400pt]{3.854pt}{0.800pt}}
\multiput(487.00,38.38)(2.271,0.560){3}{\rule{2.760pt}{0.135pt}}
\multiput(487.00,35.34)(10.271,5.000){2}{\rule{1.380pt}{0.800pt}}
\multiput(503.00,40.07)(1.690,-0.536){5}{\rule{2.467pt}{0.129pt}}
\multiput(503.00,40.34)(11.880,-6.000){2}{\rule{1.233pt}{0.800pt}}
\multiput(521.41,36.00)(0.507,0.824){25}{\rule{0.122pt}{1.500pt}}
\multiput(518.34,36.00)(16.000,22.887){2}{\rule{0.800pt}{0.750pt}}
\multiput(537.41,56.87)(0.507,-0.649){27}{\rule{0.122pt}{1.235pt}}
\multiput(534.34,59.44)(17.000,-19.436){2}{\rule{0.800pt}{0.618pt}}
\multiput(553.00,41.40)(0.739,0.512){15}{\rule{1.364pt}{0.123pt}}
\multiput(553.00,38.34)(13.170,11.000){2}{\rule{0.682pt}{0.800pt}}
\multiput(570.41,51.00)(0.507,0.556){27}{\rule{0.122pt}{1.094pt}}
\multiput(567.34,51.00)(17.000,16.729){2}{\rule{0.800pt}{0.547pt}}
\multiput(587.41,70.00)(0.507,2.073){27}{\rule{0.122pt}{3.400pt}}
\multiput(584.34,70.00)(17.000,60.943){2}{\rule{0.800pt}{1.700pt}}
\multiput(604.41,138.00)(0.507,2.904){25}{\rule{0.122pt}{4.650pt}}
\multiput(601.34,138.00)(16.000,79.349){2}{\rule{0.800pt}{2.325pt}}
\multiput(620.41,227.00)(0.507,1.546){27}{\rule{0.122pt}{2.600pt}}
\multiput(617.34,227.00)(17.000,45.604){2}{\rule{0.800pt}{1.300pt}}
\multiput(636.00,279.41)(0.494,0.507){25}{\rule{1.000pt}{0.122pt}}
\multiput(636.00,276.34)(13.924,16.000){2}{\rule{0.500pt}{0.800pt}}
\multiput(653.41,286.53)(0.507,-1.020){27}{\rule{0.122pt}{1.800pt}}
\multiput(650.34,290.26)(17.000,-30.264){2}{\rule{0.800pt}{0.900pt}}
\multiput(670.41,247.75)(0.507,-1.781){25}{\rule{0.122pt}{2.950pt}}
\multiput(667.34,253.88)(16.000,-48.877){2}{\rule{0.800pt}{1.475pt}}
\multiput(686.41,171.94)(0.507,-5.075){27}{\rule{0.122pt}{7.965pt}}
\multiput(683.34,188.47)(17.000,-148.469){2}{\rule{0.800pt}{3.982pt}}
\put(702,36.34){\rule{3.400pt}{0.800pt}}
\multiput(702.00,38.34)(8.943,-4.000){2}{\rule{1.700pt}{0.800pt}}
\put(470.0,37.0){\rule[-0.400pt]{4.095pt}{0.800pt}}
\end{picture}

\vspace{0.3cm}

{\small {\bf Fig.2} Electronic Raman continuum intensity versus
frequency ( in meV ) in the $B_{1g}$ geometry at $T= 20^o$K, for
different dopings.} \\

With larger doping the fermi surface moves away from the
region where the gap is maximum. So for the $B_{1g}$ geometry
the peak in the Raman spectra will shift towards lower frequency 
(cf Fig.2).
On the other hand for the $A_{1g}$ mode most of the
contributions are from the regions away from corners where the
SC gap is large, because the structure factor is very small there.
Therefore, for the $A_{1g}$ mode, quasiparticles from almost all
regions contribute equally to the scattered intensity and hence
for small doping variations or for small shrinkage of the Fermi
surface, there is no noticable effects at all (Fig.1). \\

\onecolumn
\begin{center}
{\bf FIGURE CAPTIONS}
\end{center}

\noindent {\bf Fig.1} Electronic Raman continuum intensity versus
frequency ( in meV ) in the $A_{1g}$ geometry at T= $20^o$K,
for different dopings. \\

\noindent {\bf Fig.2} Electronic Raman continuum intensity versus
frequency ( in meV ) in the $B_{1g}$ geometry at $T= 20^o$K, for
different dopings. \\

\begin{thebibliography}{99}
\bibitem{1} X. K. Chen, E. Altendorf, J. C. Arwin, R. Liang and
H. Hardy, Phys. Rev. {\bf B 48} (1993), 10530 ; Phys. Rev. {\bf
B 47 } (1993) 8140.
\bibitem{2} S. Sugai, Y. Enomoto and T. Muramaki, Solid. State.
Commn {\bf 75} (1990), 975.
\bibitem{3} S. L. Cooper, F. Slakey, M. V. Klein, J. P. Rice and
D. M. Ginsberg, Phys. Rev. {\bf B 37} (1988), 5920 ; 
Phys. Rev. {\bf B 38} (1988), 11934. 
\bibitem{4} T. Staufer, R. Newetchek, R. hackl. P. Muller and H.
Veith, Phys. Rev. Lett. {\bf 68} (1992), 1069. 
\bibitem{dev} T. Devereaux, D. Einzel, B. Stadtlober, R. Hackl,
D. H. Leach and J. J. Neumeier, Phys. Rev. Lett. {\bf 72}
(1994), 396.
\bibitem{screen} M. V. Klein and S. B. Dierker, Phys. Rev {\bf B
29} (1983), 4976. 
\bibitem{8} X. K. Chen, J. C. Irwin, R. Liang and W. N. Hardy,
Physica {\bf C 227} (1994), 113.
\bibitem{9} J. R. Schrieffer, X. G. Wen and S. C. Zhang, Phys.
Rev B 39 (1989), 11663 ; A. P. Kampf and J. R. Schrieffer, 
 Phys. Rev. {\bf B 41} (1990), 6399.
\bibitem{10} H. N. Ghosh and S. N. Behera, Ind. J. Phys. {\bf A
69}, (1995) 14 ; S. N. Behera and H. N. Ghosh, Z. Phys. {\bf B 95 }
(1994), 275; Haranath Ghosh and M. Sardar, Physica {\bf C 246} (1995),335. 
\bibitem{chak} S. Chakraborty, A. Sudbo, P. W. Anderson and S.
Strong, Science {\bf 261} (1993), 337. 
\end{thebibliography}
\end{document}